\documentclass[12pt,a4paper]{article}
\usepackage{jheppub}
\usepackage[T1]{fontenc}
\usepackage{slashed}

\usepackage{comment}
\usepackage{xcolor}
\usepackage{color,soul}
\usepackage{bbold}
\usepackage{graphicx}
\usepackage{feynmp}

\everymath{\displaystyle}
\usepackage{breqn}

\usepackage{amsmath}

\renewcommand{\Bbb}{\mathbb}

\usepackage{draft}

\newcommand{\h}[3]{H^#1.H^#2.H^#3}

\def\[{\left[}
\def\]{\right]}
\def\({\left(}
\def\){\right)}

\def\Y6{\text{Y}_6}
\def\B4{\text{B}_4}
\def\E2{\text{E}_2}
\def\C2{\text{C}_2}

\title{Black Hole Entropy for M-theory on the Quintic Threefold via F-theoretic Strings}

\author[1]{Indranil
  Halder\note{ihalder@g.harvard.edu},}  \author[2]{Cumrun Vafa\note{vafa@g.harvard.edu},}  \author[3]{Kai Xu\note{kaixu@math.harvard.edu}}
\affiliation[]{Department of Physics, Harvard University, Cambridge, MA 02138, USA}
\abstract{ Microscopic black hole entropy calculations in string theory usually proceeds through identifying them as wrapped strings in one higher dimension.  For M-theory on elliptic Calabi-Yau threefolds this proceeds via its relation to F-theory in one higher dimension.
Here we show how this method can be extended to M-theory on non-elliptic Calabi-Yau threefolds such as the quintic via conifold transition to elliptic threefolds.  This leads to the computation of the black hole entropy through elliptic genera of the strings.  However the Cardy formula for the computation of the black hole entropy of these strings fails because the relevant momentum excitations on the string are much smaller than the central charge of the strings.  We show how the black hole attractor entropy formula leads to predicting corrections to the Cardy formula in this regime.
}

\begin{document}

\maketitle

\section{Introduction}
In this paper we consider M theory compactified on a Calabi-Yau threefold $Y$. We are interested in charged supersymmetric black holes.  These are labeled by an electric charge $q$ which is an element of $q \in H_2(Y,\mathbb{Z})$.  In the M-theory context it arises by M2 branes wrapping this cycle.  The simplest version of such a Calabi-Yau is $T^6$ or $K3\times T^2$, and in these cases the study of such BPS black holes in 5 dimensions was one of the first successful calculations of Bekenstein-Hawking entropy from a microscopic description \cite{Strominger:1996sh}.  These also have a spinning version which in these cases was also studied \cite{Breckenridge:1996is}.  For a generic Calabi-Yau it was shown in \cite{ Gopakumar:1998ii, Gopakumar:1998jq,Katz:1999xq} that topological string theory captures the microscopic degeneracies of spinning BPS states.  There is a macroscopic prediction for large charge limit for how this entropy should behave using supergravity description \cite{Ferrara:1995ih,Ferrara:1996um}.  However the large charge limit of topological string is hard to compute (however see \cite{Huang:2007sb,Halder:2023kza} for progress in this direction), and thus the supergravity prediction is hard to verify from the microscopic picture.  For elliptic Calabi-Yau threefolds there is a way to compute the BPS black hole entropy by relating it to elliptic genus of strings in the 6d theory obtained by compactifying F-theory on the corresponding elliptic CY 3-fold \cite{Vafa:1997gr,Haghighat:2015ega}.  However this does not cover the case of non-elliptic Calabi-Yau threefolds.  The aim of this paper is to improve the situation for these cases.  

The basic strategy is to relate such manifolds to elliptic Calabi-Yau manifolds via conifold type transitions \cite{Strominger:1995cz}.  Conifold transition involves going to singular limits of the manifold and then smoothing it by deforming away by giving expectation value to some fields.  In particular we choose a set of charges in such a way that the attractor value of the Kahler class of the elliptic 3-folds lands us on a singular elliptic Calabi-Yau, which can be deformed by giving vev to some charged hypermultiplets (which correspond to complex structure deformation of the Calabi-Yau) to the smooth non-elliptic 3-fold of interest.  Since the BPS index for the black hole does not depend on the hypermultiplet moduli this does not change the resulting entropy.  In other words, the BPS index of the black hole in the non-elliptic situation can be identified with that of an elliptic 3-fold with a singular attractor value for Kahler classes.  This leads to identifying the leading large charge behavior of the BH entropy to the elliptic genus of a particular string.  One can extend this to arbitrary charges, not necessarily large, by noting that the Higgsing does not change the BPS degeneracy of any of the charged states.  This leads to considering a collection of all strings on the elliptic 3-fold side which lead to a specific charge after the geometric transition.  Summing over the corresponding elliptic genera of all these strings leads to the exact microscopic black hole entropy index of the non-elliptic threefold.  As usual we study not the entropy but the entropy index (which is convenient in the supersymmetric cases defined by a suitable supersymmetric index), and the two are expected to be close.  In this paper we focus on the quintic CY 3-fold as a prototypical non-elliptic 3-fold, but this idea should work for more general Calabi-Yau 3-folds.

Performing this study we find one surprise, which in a sense may have been anticipated:  The attractor value for the elliptic 3-fold corresponding to the point of transition to the non-elliptic 3-fold is far from the 6d limit, where the fiber size goes to zero.  Indeed the base and fiber Kahler classes are comparable at attractor values that would land one on the verge of transition to the quintic.  This in particular means we would be interested in the elliptic genus level which scales as the square root of the central charge of the string.  In this limit one cannot trust the Cardy formula for the asymptotic growth of states.   Of course if we knew the exact elliptic genus that would enable us to compute the index directly without appealing to the Cardy formula.  However such explicit descriptions are not generally available.  We show that Cardy formula does not yield the correct result for large charges.  Turning this around, using the supergravity prediction for the entropy, we can compute how the Cardy formula gets corrected in such cases!  

The organization of the rest of this paper is as follows.  In section 2 we review 5 dimensional charged BPS black holes.  In section 3 we show how F-theoretic strings on elliptic CY 3-folds can be used to compute 5d black hole entropy for M-theory comapactification on Elliptic 3-fold.  In section 4 we explain how the quintic 3-fold can make transition to genus 1 and elliptic 3-fold CY's.  In section 5 we show how to use this to relate the black hole entropy of quintic to elliptic genus of strings on the corresponding elliptic/genus one 3-folds.  We also show how the Cardy prediction gets corrected in the limit of large charges.  We also show how one goes beyond the large charge limit, and relate the entropy of black holes for arbitrary charge to a sum over elliptic genera of certain strings.

\section{5 dimensional BPS black holes}

In this section, we consider M-theory compactified on a Calabi-Yau threefold $Y$. We are interested in charged supersymmetric black holes labeled by an electric charge $q \in H_2(Y,\mathbb{Z})$.   These correspond to M2 branes wrapping the class $q$. There is a macroscopic supergravity formula predicting the entropy of such a black hole in the limit of large charges.  This is done by noting that the near horizon geometry of such BPS black holes fixes the Kahler moduli of the Calabi-Yau to minimize the corresponding mass (in Einstein frame) \cite{Ferrara:1995ih, Ferrara:1996um}.  More precisely the entropy $S$ is given by

\begin{equation}
        S= \frac{2\pi}{3} \sqrt{2 M^3};\qquad min_{k} M\qquad {\rm where}\quad M=k\cdot q, \quad   k^3=1
\end{equation}

Here $k$ is the Kahler class of $Y$.  We expect this formula to be valid whenever all the components of $q$ is large compared to one which can be conveniently achieved by $q\rightarrow N\ q$ for $N \gg 1$.  The minimization problem can be solved explicitly in terms of a rescaled Kahler class $\hat k=M^{\frac{1}{2}} k$ leading to
\begin{equation}\label{attractor2}
    \begin{aligned}
        S=\frac{ 2\pi \sqrt{2}}{3} \  {\hat k}^3, \quad {\hat k}^2=q
    \end{aligned}
\end{equation}
Below for simplicity of notation we denote this rescaled Kahler class by $k$.  We are interested in obtaining this entropy from the microscopic perspective, and in particular from the perspective of a wrapped string in one higher dimension.
We will focus on the prototype example of quintic threefold and try to understand the microsopic origin of such a formula from the perspective of BPS strings in one higher dimension. Before we proceed to a complicated example such as quintic, in the next section we review (and extend) the results in \cite{Vafa:1997gr,Haghighat:2015ega} for compactification of M-theory on an elliptic Calabi-Yau threefold.  In such cases the microscopic origin of BPS black holes can be traced to elliptic genus of strings in one higher dimension.  Moreover when the charge in the elliptic fibre (which is the same as the momentum of wrapped string) is large  compared to that on the base we get an effective Cardy type formula for BPS black hole entropy from F-theoretic string in 6 dimensions on the coresponding elliptic Calabi-Yau.  We also show that even though this microscopic description always works the Cardy formula is not valid if the charge in the elliptic fiber direction is not large enough.  This is expected from the 2d perspective which requires large energies for the Cardy formula to be applicable. We shall see that the supergravity is also compatible with this break down of Cardy formula and predicts the corrections to it!

\section{Genus 1, elliptic CY$_3$ and F-theoretic strings}\label{sec3}

In this section we focus on a compact genus 1 Calabi-Yau threefold $Y$.  The  elliptic fiber and the base would be denoted as $E$ and  $B$ respectively. Consider F-theory compactified on $Y$ - this is the same as IIB on $B$ 
with $7$-branes on the discriminant loci and the IIB coupling constant is determined by the complex structure of the elliptic fiber $Y \to B$.   D3 branes  wrapping on a holomorphic two cycle $C \subset B$ gives us a string in non-compact six dimensional spacetime.  We will call such a string an F-theoretic string.  If the corresponding genus 1 fibration has a section, mathematically it is called an elliptic 3-fold. The section might not be unique, but form a group called Mordell-Weil group, whose free part gives rise to 6d $U(1)$ gauge symmetry and torsion part determines the global structure of the gauge group \cite{morrison2012f}.  Typically what could happen is that the genus one fibration does not have an unambiguous section, but has multi-sections which permute as we go around the base of the fibration. In such a case from the 6 dimensional perspective this action on the multi-section of the genus one fibration corresponds to having a discrete gauge group \cite{braun2014f} which is called the Tate-Shaferevich (TS) group \cite{dolgachev1992elliptic}.

We further, compactify F-theory on $S^1$ with suitable Wilson line determined by the class $[Y]$ in the TS group (when this class is trivial it would correspond to M-theory on a 3-fold with a section, i.e. an elliptic 3-fold). This gives us M-theory on $Y$. D3 branes mentioned previously, once wrapped around the circle\footnote{Wrapping it $k$ times around the circle would correspond to M2 branes wrapping $k\ C$.} maps to M2 branes wrapping $C$. In addition we get M2 wrapped on  $nE$ if we look for states on the string with $n$ units of  momentum on $S^1$. These M2 branes in five non-compact directions give point like excitations suitable for being the microstates of BPS black holes.  Instead of entropy, we will focus on entropy index (which in the leading order is expected to be equal to the entropy), which counts the net BPS states (as defined in \cite{Gopakumar:1998ii,Gopakumar:1998jq}).  From the string perspective this correspond to the elliptic genus of the F-theoretic string \cite{Vafa:1997gr,Haghighat:2013gba} which in the large momentum limit $n\gg C$ is given by 
\begin{equation}\label{Fstring}
    S  \underset{n\gg C}{\approx} 2 \pi \sqrt{n\(\frac{1}{2}C^2+1+\frac{3}{2}c_1\cdot C\)} \underset{n\gg C\gg1 }{\approx}   2\pi \sqrt{\frac{n}{2}C^2}.
\end{equation} 

The above formula is not the most general charge configuation for M-theory on elliptic threefolds.  The reason is that some charge classes in 5d come, not from D3 branes wrapping a cycle in the base of the F-theory or the momentum, but correspond to an excitation on the string carrying a 6d $U(1)$ charge, which come from Kahler classes in the elliptic fiber, and not in the base.  Here we generalize the above formula to this case as well.  The rank of the $U(1)$'s in F-theory is given by the Mordell-Weil (MW) rank of $Y\longrightarrow B$.  If the MW rank is $r$, there is a $U(1)^r$ gauge symmetry in the 6d theory \cite{morrison2012f}, which in general gives rise to a current algebra on the F-theoretic string.
Such a current algebra has a level given by an $r\times r$ matrix $k$ which can be identified with the `height pairing' on the Mordell-Weil group  \cite{park2012anomaly}, which we will now review.
In the M-theory picture, the M2 branes wrapped on $\Bbb{P}^1$'s in the fiber of $Y$ give the 5d picture of the charged particles, and the charges are calculated by the intersection numbers between fibral $\Bbb{P}^1$ and sections.  Any 2-cycle can be decomposed as $q=nE+ m_i C^i+ l_\alpha P^\alpha$, where we $C^i$ are two cycles in the base and $P^\alpha$ are fibral $\Bbb{P}^1$ which avoid the base. The M2 brane wrapped along $q$ corresponds to an F-theoretic string state for $C= m_i C^i\subset B$ with momentum $n$ and $U(1)$ charges $l_\alpha$.
The level matrix of the current algebra for string wrapped on $C$ is given by the (positive definite) height pairing on the MW group \cite{park2012anomaly, shimizu2016anomaly}:
\begin{equation}
 k_{\alpha \beta}=-S_\alpha\cdot S_\beta \cdot \hat{C}   
\end{equation}
where $S_\alpha$ are dual 4-cycles to $P^\alpha$ and $\hat{C}$ is $C$ times the elliptic fiber, i.e., $\hat{C}.E=\hat{C}.P^\alpha=0, \hat{C}.C^i=m_i$.
The BPS degeneracy is also predicted by the modified Cardy formula where we subtract the dimension of the $U(1)$ charged state given by $l$ leading to a shift $n\mapsto n-l^2/(2k)$ where by $l^2/(2k)=l_\alpha l_\beta k^{\alpha\beta}/2$ where $k^{\alpha\beta}$ is the inverse of the current algebra level matrix.  As we will note in the next subsection, there is a further shift in momentum, due to induced momentum charge on the wrapped brane. Furthermore, we extend this discussion to genus 1 threefolds. 

Now we turn to explain when and how this expectation is fulfilled from the macroscopic supergravity perspective in 5 dimensions.

\subsection{The Cardy limit of the attractor equations}

Let us first consider an elliptic Calabi-Yau threefold $Y\overset{\pi}{\longrightarrow} B$, we now show that in the limit when the charge associated with the elliptic fiber $n$ is large compared to other components of $d$ the entropy of a non-rotating BPS black hole is given by the Cardy formula in (\ref{Fstring}). 

We can choose a basis $C^i$ of $H_2(B)$ with dual basis $C^i$ under intersection pairing on $B$, and also choose a basis of $H_4(X)$ given by $\hat{C}_i=\pi^{-1}(C^i)$ and sections $S_s,S_\alpha$ of $\pi$.  Here $S_s$ denotes the canonical section defining the elliptic 3-fold and $S_\alpha$'s denote the other sections.  We denote the corresponding (Poincare dual) Kahler class by $k^i,k^s,k^\alpha$, and a general Kahler class can be expressed as $k= t_ik^i+t_sk^s+ t_\alpha k^\alpha$ We may seperate it into base and fiber directions $k_b= t_ik^i$ and $k_f=t_sk^s+ t_\alpha k^\alpha$.\footnote{Note that this decomposition is not canonical and depend on the choice of the basis elements $k^s,k^\alpha$.} Then the volume $V=k^3=3k_b^2k_f+3k_bk_f^2+k_f^3$. One may expand this and get a cubic polynomial in $t_i,t_s,t_\alpha$ with coefficients given by triple intersections. We are interested in the elliptic limit $k_b \gg k_f$, i.e. $t_i \gg t_s,t_\alpha$, and the leading terms in this limit is $3k_b^2k_f=3t_ft_it_jd_{ij}$ where $t_f=t_s+\sum t_\alpha$ and $d_{ij}=C^i\cdot C^j$ is the intersection pairing on the base $B$. The first order correction is given by $3k_bk_f^2$.

The attractor equation $k^2=q$ uniquely determines a 2-cycle $q$, and we may write $q=nE+ m_i C^i+ l_\alpha P^\alpha$, where we view $C^i$ as two cycles in $Y$ through the chosen section, and $P^\alpha$ are vertical fibers which intersect $S_\alpha$ once and avoids the rest of the sections. Then we can relate $n,m_i,l_\alpha$ to $t_s,t_i,t_\alpha$ as follows (using $k_b^2(k_f-t_fk^\alpha)=0,$ which is valid for all $\alpha$'s):
\begin{equation}\label{charges}
    \begin{aligned}
        & m_i=k_iq=k_ik^2= 2t_f t_i+k_ik_f^2\\
        & n-c_1(B). \  m_i C^i=k^sq=k^sk^2=t_it_jd_{ij}+2k^sk_bk_f+k^sk_f^2\\
        & n+ m_i k^ik^\alpha k^s+l_\alpha=k^\alpha q=k^\alpha k^2=t_it_jd_{ij}+2k^\alpha k_bk_f+k^\alpha k_f^2
    \end{aligned}
\end{equation}
Hence we may also solve $n,l_\alpha$ and $C^2$: 
\begin{equation}
    \begin{aligned}
        & n=t_it_jd_{ij}+2k_bk^s(k_f-t_fk^s)\\
        & l_\alpha=2(k_f-t_fk^s)(k^\alpha-k^s)k_b\\
        & C^2=( m_i C^i)^2=(\sum 2t_f k_b+k_f^2E )^2=4t_fk_b^2k_f+4t_fk_bk_f^2+(k_f^2E)^2
    \end{aligned}
\end{equation}
Now we may compare the entropy obtained from the attractor mechanism
\begin{equation}
    S=\frac{2\pi\sqrt{2}}{3}k^3=\frac{2\pi\sqrt{2}}{3}(3k_b^2k_f+3k_bk_f^2+k_f^3)=2\pi \sqrt{2}\ k^2_bk_f \bigg(1+O\bigg(\frac{k_f}{k_b}\bigg)\bigg)
\end{equation}
with the Cardy formula expected from the F theoretic point of view
\begin{equation}
    S_{Cardy}=2\pi\sqrt{\frac{1}{2}nC^2}=2\pi \sqrt{2}\ k^2_bk_f \bigg(1+O\bigg(\frac{k_f}{k_b}\bigg)\bigg)
\end{equation}
We see that in the elliptic limit $k_b \gg k_f$ both of these coincide as expected. 

Now we compute the first-order correction to the leading order Cardy formula above.
To this end we note that the level of $U(1)$ current algebra is given by the height pairing
\begin{equation}
    \begin{aligned}
        & k_{\alpha\beta}=-(k^\alpha-k^s)(k^\beta-k^s)( m_ik^i)\\
    \end{aligned}
\end{equation}
The Sugawara dimension of the $U(1)$ current is given by
\begin{equation}
    \begin{aligned}
        & \frac{l^2}{2k} \equiv \frac{l_\alpha k^{\alpha\beta}l_\beta}{2}=-\frac{1}{t_f}(k_f-t_fk^s)^2k_b\bigg(1+O\bigg(\frac{k_f}{k_b}\bigg)\bigg)
    \end{aligned}
\end{equation}
Up to $O(k_b)$ the momentum on the elliptic fibre and the central charge of the conformal field theory on the string is given by
\begin{equation}
    \begin{aligned}
        & n=k_b^2k^s+2k_bk^s(k_f-t_fk^s)+O(k_b^0)\\
        & C^2=4t_fk_b^2k_f+4t_fk_bk_f^2+O(k_b^0)
    \end{aligned}
\end{equation}
The attractor value of the entropy can be written as follows\footnote{
Note that 
$$-\frac{l_\alpha k^{\alpha\beta}l_\beta}{c_1(B)C}=-\frac{(k_f/t_f-k^s)^2k_b}{(k^s)^2k_b}\bigg(1+O\bigg(\frac{k_f}{k_b}\bigg)\bigg)$$
is bounded by the norm of the height pairing, hence the $l^2/(2k)$ term is never larger than $c_1(B)\cdot C/2$ unless attractor flow crosses a domain wall which flops the Calabi-Yau.}
\begin{equation}
    \begin{aligned}
        S&=2\pi\sqrt{2} (k^2_bk_f+k_bk_f^2)+O(k_b^0)\\
        &=2\pi\sqrt{2((k^2_bk_f)^2+2(k^2_bk_f)(k_bk_f^2))}+O(k_b^0)\\
        &=2\pi\sqrt{\frac{1}{2}\bigg(n-\frac{l^2}{2k}-\frac{1}{2}c_1(B)\cdot C\bigg)C^2}+O\bigg(\frac{C^2}{n}\bigg)
    \end{aligned}
\end{equation}
In the last line we used the second equation in (\ref{charges}). We emphasize that we have derived the modified version of the Cardy formula
\begin{equation}
    S_{Cardy'}=2\pi\sqrt{\frac{1}{2}\bigg(n-\frac{l^2}{2k}-\frac{1}{2}c_1(B)\cdot C\bigg)C^2}
\end{equation}
from leading order super-gravity attractor mechanism in the limit $k_b \gg k_f$, i.e., $n\gg l^2/k,C\gg 1$. 
This is precisely what one would expect from a Cardy type formula, namely, we subtract $l^2/(2k)$ which is the dimension of the state with charge $l$ given by current algebra level matrix $k$ from the momentum.  Moreover the extra shift by $-c_1(B)\cdot C/2$ signifies the fact that a D3 brane wrapped around $C\times S^1$ carries extra momentum along $S^1$.  To see this, note that in the M-theory (in one lower dimension) this gets mapped to an M5 brane wrapped on the elliptic fibration over $C$ times a circle, which we denote by $\hat C\times S^1$.  Viewing $S^1$ as the 11-th dimension, in the IIA picture this is the same as D4 brane wrapping $\hat C$ and the momentum gets mapped to D0 brane charge.  Since wrapped D4 brane on $\hat C$ carries D0
brane charge \cite{vafa1996gas} given by  $-c_2(\hat C)/24=-c_1(B) \cdot C/2$, we learn that D3 brane wrapping $C \times S^1$ receives such an extra momentum shift.

We may also directly observe that if the volume $V$ factorizes into a product of the base and fiber exactly, in this case we have $k_bk^\alpha k^\beta=k_bk^\alpha k^s=k_b(k^s)^2=0$. Then all the formulas above get significantly simplified:
$$C=2t_f k_b,\,n=k_b^2,\, V=k_b^2k_f$$
And we can see that $S=2\pi \sqrt{2} k^2_bk_f=S_{Cardy}$ exactly.  So we see that whether there are strong corrections to Cardy formula in the leading large charge limit or not, would depend on whether the volume of the elliptic 3-fold which makes a transition to the quintic has a product structure or not.

\subsection{Extension to genus 1 fibrations}

F-theory requires only to have a well defined $\tau$ (complex moduli of the elliptic fiber) up to modular transformations, as a function of base.  However, it does not require having a section for the elliptic fibration.  In such cases CY will have multi-sections and are called genus 1 fibration.
In such cases the physical theory in 6 dimensions will have a discrete gauge symmetry \cite{braun2014f} and as we go down to 5 dimensions we can turn on a non-trivial element $g$ of the holonomy of this discrete gauge symmetry\footnote{Not turning on this holonomy would give rise to an elliptic threefold with section. So multiple 5d theories, given by different CY 3-folds are related to the same 6d theory, and are distinguished only by a discrete holonomy around the circle compactification to 5d.}.

For a genus 1 fibration $Y\longrightarrow B$ we have an associated elliptic fibration $J(Y)\longrightarrow B$ to which the argument in the previous section applies. Now we can fix a multi-section $\Sigma_s$ of degree $m$ and then obtain a basis of $H_4$ given by $\hat{C}_i,\Sigma_s,\Sigma_\alpha=S_\alpha(\Sigma_s)$, where $S_\alpha$ are sections of the elliptic Calabi-Yau $J(Y)$ as we have chosen in the last section, and by $S_\alpha(\Sigma_s)$ we mean the image of the 4-cycle $\Sigma_s$ in $Y$ under the action\footnote{Fibers of $J(Y)$ act on the corresponding fibers of $Y$, hence sections of $J(Y)$ act on the total space of $Y$. In other words, we have a natural action of $MW(J(Y))$ on $Y$.} of $S_\alpha$, which is another 4-cycle in $Y$ \cite{morrison2012f}. Here we use the fiberwise action of $J(Y)$ on $Y$. Then we have a similar decomposition of Kahler classes $k=t_ik^i+t_sk^s+ t_\alpha k^\alpha$, where $k^\alpha$ is the Poincare dual of $\Sigma_\alpha/m$. We normalize in this way so that adding $k^\alpha$ increases the volume of elliptic fiber $E$ by $1$. Then the same argument as for the elliptic 3-fold works, except for the fact that  $(k^s)^2k^i=c_1(\Sigma_s)\cdot C/m$ instead of $c_1(B)\cdot C$. Another difference, which does not affect the calculation is that the 6d string which wraps the circle experiences the discrete gauge symmetry $g$ along the twisted circle by being in the sector ${\cal H}_g$ twisted by $g$, as in orbifold constructions, and so the elliptic genus of the string is calculated in this twisted sector.

{
\section{From quintic to a genus 1 and elliptic CY$_3$}

In this section we construct two geometric transitions to the quintic CY 3-fold.  One from a genus one CY 3-fold and another from an elliptic CY 3-fold.
\subsection{Singular limit of the quintic and genus 1 CY 3-fold}

Here we give a conifold transition from quintic $Y$ to a genus one fibered manifold $Y_2$ (i.e. it is fibered by torus but does not have a section).

A smooth quintic is defined by a generic polynomial $P_5(x)$ in $x_i\,(i=1,\cdots 5)$ of degree $5$. We may tune the complex structure (i.e. the coefficients of $P_5$) to the following special form satisfying
\begin{equation}
    P_5(x)=\det \begin{pmatrix}
u^1_1(x) & u^2_1(x) & u^3_1(x) \\
v^1_1(x) & v^2_1(x) & v^3_1(x) \\
q^1_3(x) & q^2_3(x) & q^3_3(x)
\end{pmatrix}=\sum \epsilon_{abc}u^a_1(x)v^b_1(x)q^c_3(x)=0
\end{equation}
where $u^a_1(x),v^b_1(x),q^c_3(x)$ are polynomials in $x_i$ of degree $1,1,3$ (as indicated by the subscript). For example, we may pick
\begin{equation}
  P_5(x)=\det \begin{pmatrix}
x_1 & x_2 & x_3 \\
x_4 & x_5 & x_1 \\
x_2^3 & x_3^3 & x_4^3
\end{pmatrix}=x_1x_4^3x_5+x_1x_2^4+x_3^4x_4-x_2^3x_3x_5-x_2x_4^4-x_1^2x_3^3=0  
\end{equation}
This defines a singular quintic $Y=\{[x]=[x_i]\subset\Bbb{P}^4|P_5(x)=0\}$.

A resolution $Y_2$ of $Y$, which has one additional Kähler parameter, is defined by the following equation as a subvariety in $\{[x]\times [y]\}=\Bbb{P}^4\times \Bbb{P}^2$: 
\begin{equation}\label{y2def}
\begin{aligned}
    \begin{pmatrix}
U_{11}(x,y)\\V_{11}(x,y)\\Q_{31}(x,y)
\end{pmatrix}=\begin{pmatrix}
u^1_1(x) & u^2_1(x) & u^3_1(x) \\
v^1_1(x) & v^2_1(x) & v^3_1(x) \\
q^1_3(x) & q^2_3(x) & q^3_3(x)
\end{pmatrix}\begin{pmatrix}
y^1\\y^2\\y^3
\end{pmatrix}=0
\end{aligned}
\end{equation}
Schematically this is written as defined by three equations with degrees
\begin{equation}\label{y2def2}
    \begin{aligned}
        \quad & \Bbb{P}^2 | &1 \quad &1 &1\\
       & \Bbb{P}^4 | &3 \quad &1  &1
    \end{aligned}
\end{equation}
Here each column in the matrix is associated with a polynomial in (\ref{y2def}) and the entries in the column denote the degree of the polynomial in the respective projective space. $Y_2$ has $h^{1,1}=2$ Kahler moduli and $h^{2,1}=68$ complex structure moduli. 
 The nonzero triple intersection number of four cycles are $C_{112}=3,C_{122}=7,C_{222}=5$.\footnote{This is computed by the intersection theory: $\h{1}{1}{2}$ on $Y_2$ is equal to $\h{1}{1}{2}(H^1+H^2)(H^1+4H^2)$ on $\Bbb{P}^2\times \Bbb{P}^4$. From now on we will work in this basis and denote it as  $H^1=(1,0), H^2=(0,1)$.}

We have a natural projection map $Y_2\longrightarrow Y$ by forgetting the $y$ coordinates, and the preimage of $[x]$ is a point if $$\mathrm{rank}\begin{pmatrix}
u^1_1(x) & u^2_1(x) & u^3_1(x) \\
v^1_1(x) & v^2_1(x) & v^3_1(x) \\
q^1_3(x) & q^2_3(x) & q^3_3(x)
\end{pmatrix}=2$$ (which happens generically on $Y$ if $u,v,q$ are sufficiently general), the preimage of $[x]$ is $\Bbb{P}^1$ if $$\mathrm{rank}\begin{pmatrix}
u^1_1(x) & u^2_1(x) & u^3_1(x) \\
v^1_1(x) & v^2_1(x) & v^3_1(x) \\
q^1_3(x) & q^2_3(x) & q^3_3(x)
\end{pmatrix}=1$$ the preimage of $[x]$ is $\Bbb{P}^2$ if $$\mathrm{rank}\begin{pmatrix}
u^1_1(x) & u^2_1(x) & u^3_1(x) \\
v^1_1(x) & v^2_1(x) & v^3_1(x) \\
q^1_3(x) & q^2_3(x) & q^3_3(x)
\end{pmatrix}=0$$ (which does not happens if $u,v,q$ are sufficiently general.) One finds that $Y_2\longrightarrow Y$ is a blow up of 34 different $\Bbb{P}^1$'s within the same class of $H_2$.

Now we have another natural map $Y_2\longrightarrow \Bbb{P}^2$ by forgetting the $[x]$ coordinate. The fibers of this map are elliptic curves. This is because if we fix the Kahler class of the fiber and let the base class goes to infinity, $Y_2$ would converge to $fiber\times \mathbb{R}^4$, for this to be a Calabi-Yau fiber must be an elliptic curve. We can also see this concretely: for each $y$, we may solve $x_4,x_5$ as linear combinations of $x_1,x_2,x_3$ by $U_{11}(x,y)=V_{11}(x,y)=0$. Then $Q_{31}(x,y)=0$ becomes a cubic equation in $x_1,x_2,x_3$ with coefficients determined by $y$. In other words we have a family of cubic curves parameterized by $[y]\in \Bbb{P}^2$, which gives a genus 1 fibration. We may choose base and fiber 2-cycles $H=(1,0)$ and $E=(0,3)$. The TS group for $Y_2$ is $\mathbb{Z}_3$, leading to this discrete gauge symmetry in 6d.

\subsection{Singular limit of the quintic and elliptic CY 3-fold}
The genus one Calabi-Yau $Y_2$ has no elliptic sections (it has tri-section).  We can consider a more singular locus of quintic if we want to relate it to an elliptic CY 3-fold (by elliptic Calabi-Yau threefold it is meant in the math literature that it has a section).
If we replace the rank 3 matrix in the previous section by a rank 2 matrix, we would get a transition from the quintic to a K3 fibration $Y_1\longrightarrow \Bbb{P}^1$. The transition point is defined by the singular quintic equation
\begin{equation}
    P_5(x)=\det \begin{pmatrix}
l^1_1(x) & l^2_1(x) \\
q^1_4(x) & q^2_4(x) 
\end{pmatrix}=\sum \epsilon_{\alpha\beta}l^\alpha_1(x)q^\beta_4(x)=0
\end{equation}
We may combine these two transitions to get a transition from $Y$ to $Y_{12}$, which admits both a genus 1 fibration and a K3 fibration. These two fibrations are transverse, i.e. the torus fibers intersect K3 fibers at discrete points. The transition point is obtained by combining the singularities of $\sum \epsilon_{\alpha\beta}l^\alpha_1(x)q^\beta_4(x)$ and $\sum \epsilon_{abc}u^a_1(x)v^b_1(x)q^c_3(x)$ into 
\begin{equation}
    P_5(x)=\sum \epsilon_{abc}\epsilon_{\alpha\beta}l^\alpha_1(x) u^a_1(x)v^b_1(x)q^{c\beta}_2(x)=0
\end{equation}
which defines $Y_{12}\subset \Bbb{P}^1\times \Bbb{P}^2\times \Bbb{P}^4$ schematically denoted as
\begin{equation}
    \begin{aligned}
       & \Bbb{P}^2 | &1 \quad &1 \quad 1 \quad 0\\
       & \Bbb{P}^1 | &1 \quad &0 \quad 0 \quad 1\\
       & \Bbb{P}^4 | &2 \quad &1 \quad 1 \quad 1
    \end{aligned}
\end{equation}
It has two extra Kähler parameters compared to quintic (leading to $(h^{1,1},h^{2,1})=(3,56)$).
There is an extra bonus of this combination: we get a section of the genus one fibration defined by $l^\alpha_1(x)=0$. This defines a linear $\Bbb{P}^2\subset \Bbb{P}^4$ which is contained in $Y$, and its projection to the base $\Bbb{P}^2$ is bijective. The preimage of $[y]$ is defined by four linear equations
\begin{equation}
    U_{11}(x,y)=V_{11}(x,y)=l_1^1(x)=l_1^2(x)=0
\end{equation}
hence consists of a single point.  Thus this is not a multi-section, and it defines a consistent section of the elliptic Calabi-Yau 3-fold. Similar to the case $Y_2$, we have $H_2(Y_{12})\simeq H_2(\Bbb{P}^2\times \Bbb{P}^1\times \Bbb{P}^4)\simeq \mathbb{Z}^3$ and $H^2(Y_{12})\simeq H^2(\Bbb{P}^2\times \Bbb{P}^1\times \Bbb{P}^4)\simeq \mathbb{Z}^3$. The nonzero triple intersection numbers are $C_{112}=2,C_{113}=3,C_{123}=5,C_{133}=7,C_{233}=4,C_{333}=5$, and the 2-cycles on the base, fiber and charged under $U(1)$ are $H=(1,5,2),E=(0,2,3),P=(0,1,1)$.

\section{Interpolation from Cardy to quintic limit}\label{4.2}
Here we focus on finding the microscopic picture for BPS black holes for the quintic. The basic idea is to choose BPS black hole charges for genus 1 or elliptic 3-fold so that the attractor point is at the transition point to the quintic.  We then use the fact that changing the complex structure to smooth out the manifold to the quintic (Higgsing some $U(1)$'s) does not change the black hole entropy.  Therefore the black hole entropy of the quintic black hole is computed using that of the genus 1 or elliptic 3-fold, which in turn is related to the microscopic picture of elliptic genus of the corresponding 6d string. 

This is the picture in the large charge regime, where the attractor point is a good approximation.
More generally, for arbitrary charge, we find that we still can relate the BPS black hole entropy of the quintic not to one particular BPS black hole charge before transition, but to all charges which after Higgsing still lead to the relevant quintic BPS black hole charge.  In other words, the elliptic genus of a collection of 6d strings captures the BPS entropy of the quintic 3-fold black holes.  

In this section we first discuss the transition from genus one 3-fold $Y_2$ and then the elliptic one $Y_{12}$.  Finally we discuss the more general case of arbitrary charges.

\subsection{Interpolation from genus 1 to quintic}

The quintic point in the Kahler moduli of $Y_2$ is $k_Q=(0,1)$ (note that $k_Q^3=5$) and is the attractor point for the 2-cycle given by $k_Q^2=7H+(5/3)E$, in other words the charges corresponding to $(m,n)=(7,5/3)$ (up to overall rescaling). Note here we have a fractional coefficient because the momentum gets fractionalized after we turn on the $\mathbb{Z}_3$ Wilson line. The elliptic point in the Kahler moduli is $k_E=(1,0)$ and is the attractor point for the 2-cycle given by $k_E^2=3E$. We may interpolate between quintic and elliptic points by $k=ak_Q+bk_E=(b,a)$, which is the attractor point for the two cycle 
\begin{equation}
\begin{aligned}
& k^2=(7a^2+6ab)H+\frac{1}{3}(5a^2+14ab+3b^2)E\\ 
\implies & m=7a^2+6ab, \quad n= \frac{1}{3}(5a^2+14ab+3b^2)  
\end{aligned}
\end{equation}
\begin{figure}[t]
	   \centering
	   \includegraphics[width=0.9
\textwidth]{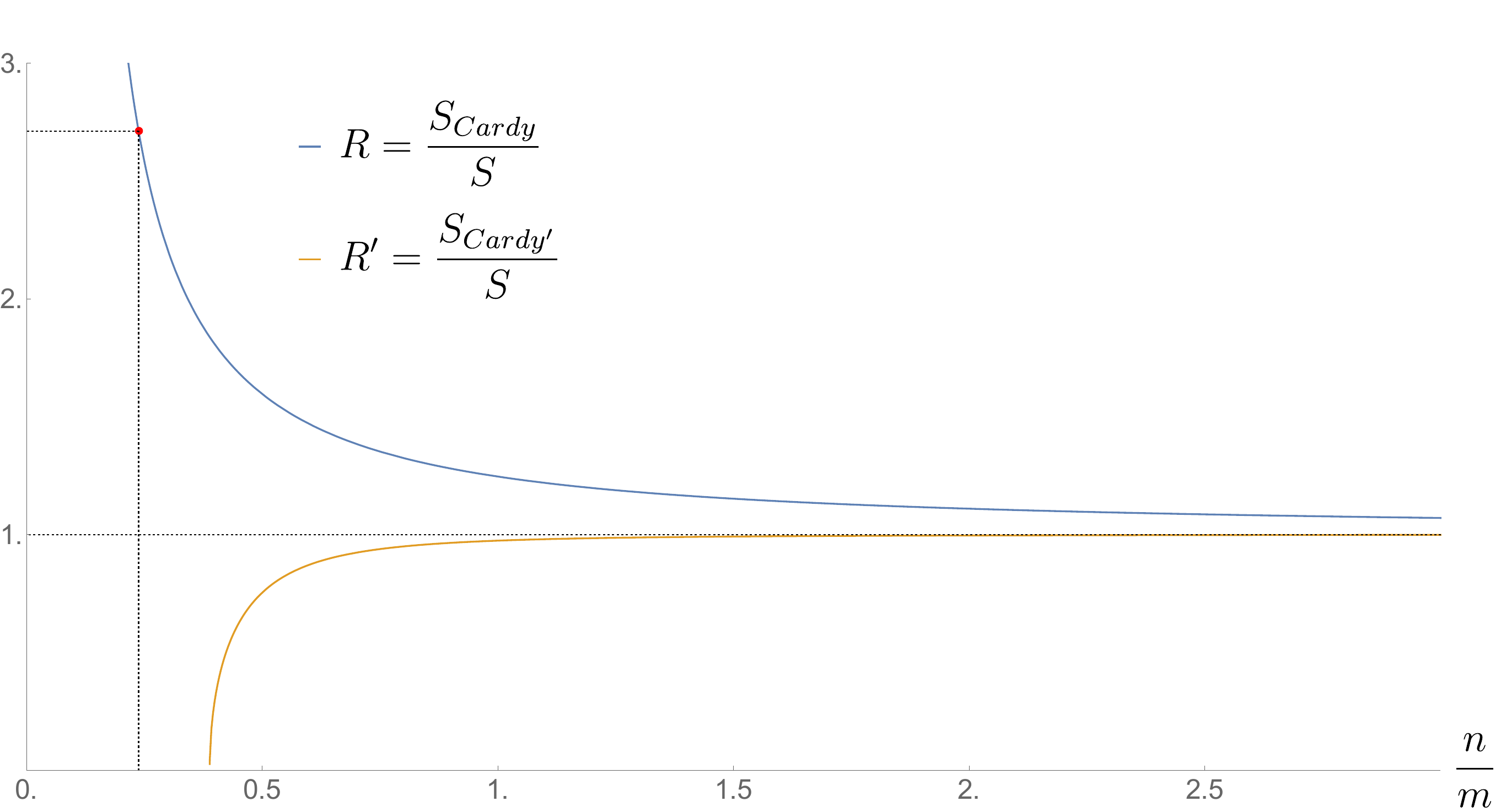}
\caption{The yellow and the blue curve represent the ratio $R'=S_{Cardy'}/S,R=S_{Cardy}/S$ between entropy of BPS blackhole as obtained from  the  Cardy formula  on $Y_2$  with and without momentum shift and the attractor value. At $n/m=5/21$ (corresponding value of $R$ is marked with a red dot) the Kahler class of $Y_{2}$ is that of quintic, whereas $n/m \gg1 $ corresponds to the situation when the dominant charge carried by the black hole is on the elliptic fibre. On the yellow curve for $R'$ as we approach the quintic point the effective shifted momentum on the elliptic fiber becomes small marking the fact that Cardy type approximation is no longer justified very close to the conifold transition.}
 \label{fig1}
\end{figure}
The attractor mechanism predicts the entropy\footnote{Given a four cycle $k$ we can identify $k^2$ with a two cycle as follows
\begin{equation}
    k^2.H^i=a^i, \quad k^2=(a^1,a^2)
\end{equation}
For instance at the quintic point, i.e.,  $b=0$, we have the following relation on quintic $q=k^2=5a^2$.
} 
\begin{equation}
    \begin{aligned}
        S&=\frac{2\pi }{3}\sqrt{2}k^3=\frac{2\pi }{3}\sqrt{2}(9ab^2+21a^2b+5a^3)
    \end{aligned}
\end{equation}
On the other hand, the leading order and modified Cardy formula give
\begin{equation}
    \begin{aligned}
        S_{Cardy}&=2\pi\sqrt{\frac{nm^2}{2}}\\
         &=\frac{\sqrt{3(5+14t+3t^2)(7+6t)^2}}{2(9t^2+21t+5)} S\\
         S_{Cardy'}&=2\pi\sqrt{\frac{m^2}{2}(n-\frac{7}{18}m)}\\
          &=\frac{\sqrt{3((5+14t+3t^2)-\frac{7}{6}(7+6t))(7+6t)^2}}{2(9t^2+21t+5)}S
    \end{aligned}
\end{equation}
where $t=b/a$ is the ratio between elliptic part of the kahler class and the quintic part, $t$ is determined by charges as follows
\begin{equation}
    \frac{n}{m}=\frac{5+14t+3t^2}{3(7+6t)}
\end{equation}
We compare $S, S_{Cardy}, S_{Cardy'}$ in figure \ref{fig1}. Note that in the elliptic limit where $n\gg m$ and $t\rightarrow \infty$ the Cardy approximation is good as we would expect.  However, as we approach the quintic point at $t=0$ (corresponding to $n/m=5/21$), the Cardy approximation breaks down, as we may have expected.

\subsection{Interpolation from elliptic to quintic}

The quintic point in the Kahler moduli is $k_Q=(0,0,1)$ (note that $k_Q^3=5$), and it is the attractor point for the 2-cycle given by $k_Q^2=(7,4,5)=7H+22E-75P$, corresponding to charges $(m,n,l)=(7,22,-75)$ (up to overall rescaling). Note here we have integral coefficients, in contrast with the genus 1 case.
The elliptic point in the Kahler moduli is $k_E=(1,0,0)$, and it is the attractor point for the 2-cycle given by $k_E^2=(0,2,3)=E$.
We may interpolate between quintic and elliptic point by $k=ak_Q+bk_E=(b,0,a)$ for some $a,b$. It is the attractor point for the 2-cycle given by $k^2=mH+nE+lP=a^2(7H+22E-75P)+2ab(3H+11E-32P)+b^2E$ where $m=7a^2+6ab,n=22a^2+22ab+b^2,l=-75a^2-64ab$.

As we noted previously, the attractor mechanism predicts the entropy 
\begin{equation}
    S=\frac{2\pi\sqrt{2}}{3}k^3=\frac{2\pi\sqrt{2}}{3}(9ab^2+21a^2b+5a^3)
\end{equation}
Moreover, the Cardy formula (with and without the momentum shift) predicts the entropy 
\begin{equation}
    \begin{aligned}
        S_{Cardy}&=2\pi\sqrt{\frac{nm^2}{2}}\\
         &=\frac{3\sqrt{(22+22t+t^2)(7+6t)^2}}{2(9t^2+21t+5)} S\\
         S_{Cardy'}&=2\pi\sqrt{\frac{m^2}{2}(n-\frac{l^2}{2k}-\frac{3}{2}m)}\\
          &=\frac{3\sqrt{(22+22t+t^2-\frac{(75+64t)^2}{64(7+6t)}-\frac{3}{2}(7+6t))(7+6t)^2}}{2(9t^2+21t+5)} S
    \end{aligned}
\end{equation}
The ratio of entropies is expressed in terms of $t=b/a$ (the ratio between elliptic part of the kahler class and the quintic part) that can be determined from
\begin{equation}
     \frac{n}{m}=\frac{22+22t+t^2}{3(7+6t)}
\end{equation}
We compare $S, S_{Cardy}, S_{Cardy'} $ in figure \ref{figE}. Note that in the large momentum limit the Cardy formula works well but as we approach the quintic point at $t=0$ (corresponding to $n/m=22/21$, Cardy formula differs significantly from the attractor value.

\begin{figure}[ht]
	   \centering
	   \includegraphics[width=0.9
\textwidth]{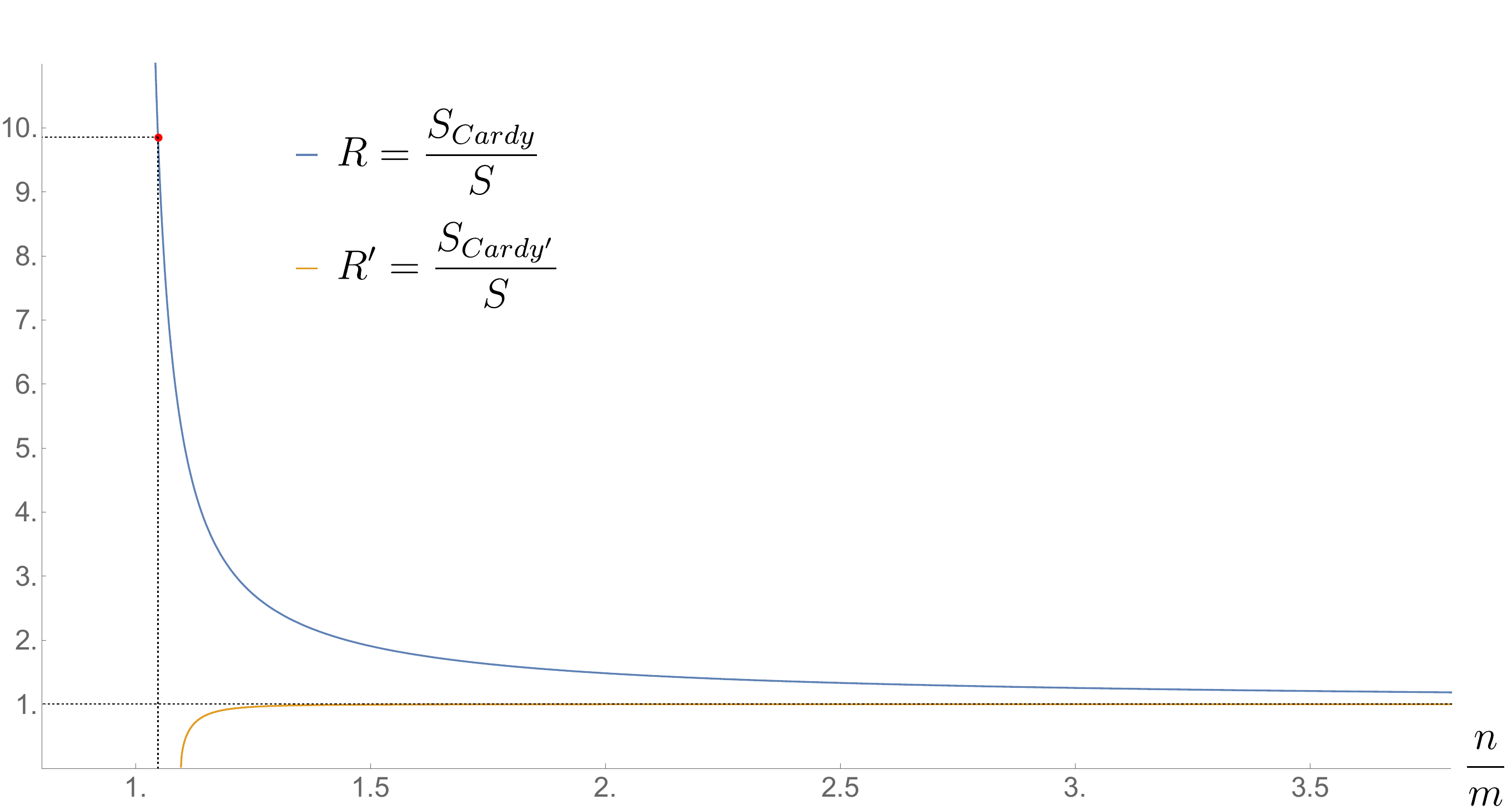}
\caption{The yellow and the blue curve represent the ratio $R'=S_{Cardy'}/S,R=S_{Cardy}/S$ between entropy of BPS blackhole as obtained from  the  Cardy formula  on $Y_{12}$  with and without momentum shift and the attractor value. At $n/m=22/21$ (corresponding value of $R$ is marked with a red dot) the Kahler class of $Y_{12}$ is that of quintic. At $n/m \gg1 $ and at the quintic limit we observe very similar behaviour to the one observed for $Y_2$.}
 \label{figE}
\end{figure}

\color{brown}

\color{black} 

\subsection{Beyond the large charge limit}

In the previous sections we have argued that in the large charge limit the elliptic genus of a string propagating in 6 dimensions can capture the index of the quintic black hole.  The question is whether this is correct beyond the leading large charge limit.  In this section we investigate this and show that this gets corrected when the charge is not large.  However, we show nevertheless that there are a collection of 6d strings whose elliptic genus sum up to capture the index of the microscopic degeneracy of the quintic.

We first show what the exact correct formula is, using the connection of the BPS states and topological strings.  And then we show how this reproduces what we found in the previous sections in the large charge limit.

Recall that the topological string partition function can be written in terms of the GV invariants obtained as M2 brane BPS index \cite{Gopakumar:1998ii, Gopakumar:1998jq}:
\begin{equation}
    Z(t,\lambda)=\prod_{q,j,-j\leq m\leq j, n} (1-e^{(2m+n)\lambda}e^{-q\cdot t})^{(-1)^{2j+1}nN^j_q}
\end{equation}
And since topological string is independent of the complex structure, $Z_{quintic}(t,\lambda)=Z_{Y_{12}}(0,0,t,\lambda)$ (and similarly for the $Y_2$ description)\footnote{This is valid up to an overall irrelevant multiplicative constant.}. This implies that 
\begin{equation}
    N^j_q=\sum_{a,b} N^j_{a,b,q}
\end{equation}
where $\lambda$ is the string coupling constant and $t$ denotes the Kahler class.  This result may also be obtained directly from geometry or from the general Higgsing picture.  Namely, giving vev to a Higgs field which does not affect some $U(1)$ gauge symmetries, does not change the spectrum of the corresponding charged states.
This in particular implies that the BPS degeneracy index of the black holes states can be computed by summing over all the states with the same charge before the transition.  In the particular cases of interest to us, this means that if we sum the elliptic genera of all the strings for the genus 1 or elliptic CY which give the same charge as the one we are interested after transition to the quintic, we get the answer for the degneracy after the transition.

Now we argue that the attractor equation indeed finds the leading term in the large charge limit for such a summation. Suppose $k_0$ is the attractor Kahler form  which satisfies the attractor equation $k_0^2=q$ with entropy $S_0=2\pi(\sqrt{2}/3) k_0^3$ and that $k_0$ is a Kahler form of the elliptic (or genus 1) 3-fold at the transition point to the qunitic.  We will now show why summing over charges with the same $U(1)$ charge of interest at the extremum point is the same as setting the other charges to 0. 
We want to show that this is an extremum point for all variations of $q$ such that $\delta q$ is part of the Higgsed $U(1)$.
Consider the attractor equation for another charge given by $q+\delta q$ in the summation, $(k_0+\delta k)^2=q+\delta q$ we obtain $2k_0. \delta k =\delta q$ where $\delta q\ll q$.  Note that $k_0\cdot \delta q=0$, because at the transition point all the charged states which are to be Higgsed have zero central charge (geometrically the correspondingly class collapses at the transition point).  So we see that $2\pi(\sqrt{2}/3) k_0^2(k_0+\delta k)=S_0+O(\delta q ^2)$
 and hence $\delta S=O(\delta q^2)$, which shows $k_0^2=q$ gives the leading contribution in the above summation.

\subsubsection*{Acknowledgments} 

The work of CV is supported by a grant from the Simons Foundation (602883,CV), the DellaPietra Foundation, and by the NSF grant PHY-2013858.  The work of IH is supported in part by DOE grants DE-SC0007870 and DE-SC0021013.

\

\providecommand{\href}[2]{#2}\begingroup\raggedright\endgroup

\end{document}